# An Underlying Theory for Gravity

**Yuan K. Ha**

Department of Physics, Temple University, Philadelphia, Pennsylvania 19122 U.S.A.

yuanha@temple.edu

**Abstract**. A new direction to understand gravity has recently been explored by considering classical gravity to be a derived interaction from an underlying theory. This underlying theory would involve new degrees of freedom at a deeper level and it would be structurally different from classical gravitation. It may conceivably be a quantum theory or a non-quantum theory. The relation between this underlying theory and Einstein's gravity is similar to the connection between statistical mechanics and thermodynamics. We discuss the apparent lack of evidence of any quantum nature of spacetime and the meaning of quantum gravity in this context.

**1. The meaning of quantum gravity**
Quantum gravity has been conjectured for almost 80 years since the introduction of the graviton. It is commonly believed that gravity is a fundamental interaction and as such, it would obey quantization similar to electrodynamics. However, it is significant to point out that there is not a single observational evidence so far showing the need of a quantum theory of gravity. On the theoretical side, despite enormous efforts in the past decades, there is still no consistent quantum gravity theory with any predictive power. The only great result from quantum gravity efforts over the last 50 years is the renormalization of Yang-Mills theory by t' Hooft [1] and Veltman [2], using the techniques developed by Feynman [3] and Dewitt [4] for perturbative calculations in general relativity.

It may be helpful to look at the problem of quantum gravity in a different perspective by asking what is *not* quantum gravity. At once, we see it is not about motion in spacetime, for motion with a precise trajectory is a classical concept. It is not about translation in spacetime. It is not about Lorentz transformation. It is not about elementary particles, or necessarily about unification of forces. Thus quantizing general relativity with special relativity as a limit is a contradiction. We recognize that the quantum gravity domain is naturally of Planck size $10^{-33}$ cm. Furthermore, that quantum gravity domain must be physical and contain real degrees of freedom, and that those degrees of freedom should also carry energy, momentum, spin and other attributes like any other excellent degrees of freedom in physics.

We should also point out that the term quantum gravity has come to acquire different meaning to different researcher. The proper case is that quantum gravity should start only with Einstein's equation and not some modification of it by adding extra terms to its Lagrangian as in modified gravity theories. We can illustrate this situation with electrodynamics. In quantum electrodynamics, the equations are exactly the same Maxwell's equations as in classical electrodynamics and not some modification of Maxwell's equations. It is well known that a remarkable modification of Maxwell's equations has become the Yang-Mills theory. Although the modified equations contain the original equations in form, the theory is completely different and it is no longer electrodynamics. Quantizing a modified theory does not lead to a successful quantum theory of electrodynamics and its spectacular experimental confirmation. Therefore current modified gravity theories are not what they intend to accomplish, namely, a quantum theory of gravitation based on general relativity. They may end up having nothing to do with ordinary gravity in the $\hbar \to 0$ limit.

**2. Quantum nature of gravity**
It is also common to find that in many quantum gravity theories there is little discussion about what constitute the quantum nature of gravity in those theories. A list of conceivable quantum effects of gravity as logical outcomes of quantum idea applied to gravity is the following:

1. Existence of graviton.
2. Spacetime fluctuations and spacetime foam.
3. Discreteness of spacetime.
4. Wave-particle duality for gravity.
5. Linear superposition of spacetimes.
6. Interference of spacetimes.
7. Uncertainty relations for spacetime metric.
8. Tunneling in gravity.
9. Entanglement in gravity.

So far none of these phenomena has been observed and the existence of any one of the above effects would be a strong indication of the validity of quantum gravity. Some of these phenomena such as superposition and interference of spacetimes are most difficult to perceive. Even the familiar graviton is only a hypothetical particle. All quantum field theories of gravity require the graviton as a mediating particle. The graviton corresponds to the weak field perturbation on a flat background spacetime. However, the graviton propagating in a flat spacetime contradicts the background independence of general relativity. Graviton theories are known to be divergent at high energies and they are not renormalizable. On the experimental side, it would be impossible to detect a *single* graviton due to its very low frequency and energy. Analysis shows that a detector must be enormous in size in order to be comparable to the graviton's wavelength. For practical observation, such a detector would become physically too massive and collapse to form a black hole, in which case the detection outcome would be lost [5]. On the other hand, an indirect evidence for the graviton would be the observation of gravitational black body radiation with a spectrum like Planck's radiation curve. Such an experimental feat would be equally daunting.

**3. Astrophysical observations**
There are at least a few efforts to search for the quantum nature of gravity in astrophysical settings. The idea of spacetime fluctuations and spacetime foam is a cornerstone prediction of quantum gravity. The observation of gamma ray bursts from a distant source could indicate a possible granular structure of spacetime at the smallest scale. A significant observation of the highest energy gamma rays has been conducted with the gamma ray burst GRB 090510 by the Fermi Gamma-Ray Space Telescope [6]. Gamma rays of energies up to 31 GeV from the burst are all seen to arrive within 1 second after traveling a distance of 7.3 billion light years. If the spread in time between the highest and the lowest-energy gamma rays is all attributed to quantum effects rather than production time difference, then the result rules out those quantum gravity theories in which the speed of light $c$ varies linearly with photon energy $E$. A variation of photon speed is an indication that Lorentz invariance is violated. A thorough analysis shows that there are no quantum effects until the distance is down to $0.8L_{Pl}$, below the Planck length. The result indicates that spacetime is classical above the Planck length and special relativity is right.

A recent investigation also set to probe the quantum nature of spacetime is carried out on observation of the farthest quasars [7]. Light from the farthest quasars with red shift $4 < z < 5.4$ are observed by the Hubble Space Telescope. The aim is to set new limits on the fluctuations of spacetime foam and quantum gravity models. An observable effect would be the degradation of the diffraction images of distant sources. Nevertheless, a careful analysis of the diffraction images of the most distant quasars shows no blurring effect that could be caused by the interaction of photons with quantum gravity fluctuations. Photons and their quantum states can travel undisturbed for billions of light years to reach the Hubble Space Telescope. The analysis takes into account the possibility of spacetime fluctuations accumulated over cosmological distances and partly compensated by the effects of phase variations. The results exclude a number of spacetime foam models based on the holographic principle. Without phase variation corrections, all major models of quantum gravity and spacetime foam are ruled out. The results also show the absence of any directional dependence of quantum gravity effects and therefore the validity of the cosmological principle by an independent method.

It is puzzling to find that these observations do not reveal any quantum nature of spacetime above the Planck length. Future detections using gamma ray bursts with even higher energy photons and observation of quasars at higher red shift will provide a more stringent limit on the validity of quantum gravity. If persistent observation cannot yield any signature of quantum spacetime, then ultimately one has to come to the conclusion that spacetime fluctuations and spacetime foam is only a preconceived notion. We have seen that parity conservation law in weak interactions was a preconceived notion without any support. The deeper question is why spacetime is so smooth and classical all the way down to the Planck length level.

**4. Higher Dimensional Theories**

A number of quantum gravity theories involve higher spacetime dimensions. Although these theories are highly appealing, they suffer in general from instability and causality problems. In Kaluza-Klein type theories of pure gravity in higher dimensions, the difficulty already exists at the classical level [8]. For point-like masses, these theories cannot produce the gravitational field which corresponds to known classical gravitational tests.

The classical tests of general relativity are the following:

1. Gravitational frequency shift.
2. Perihelion shift of Mercury
3  Deflection of light.
4. Radar echo delay
5. Parameterized post-Newtonian parameters.

A nonrelativistic weak field approximation is sufficient for the treatment of motion in the solar system. When the equations of motion are applied to the classical tests in general relativity, only the ordinary 3+1 dimensional spacetime case is found to agree with observations. The result is independent of the size of the extra dimensions as long as they are compact and have the geometry of tori. All multidimensional Kaluza-Klein theories face severe challenge when confronted with motions in the solar system.

## 5. Is gravity non-quantum?

At present, there is a total lack of evidence of any quantum nature of gravity, including spacetime foam, discrete spacetime, breaking of Lorentz invaraince, extra dimensions. Is it possible that quantum gravity is not necessary? What if gravity is intrinsically classical and it becomes a derived interaction from an underlying theory at a deeper level [9]. We presume that the underlying theory would be a completely different theory from gravitation. It may be a quantum theory or a non-quantum theory and the important point is that classical gravity is not the $\hbar \to 0$ limit of this underlying theory. The relation between classical gravitation and the underlying theory is more like that between thermodynamics and statistical mechanics. In classical thermodynamics, the basic variables are the pressure $P$, volume $V$, temperature $T$ and entropy $S$. They are macroscopic variables chosen for performing heat experiments. The underlying degrees of freedom of the system are the microscopic atoms and molecules. They obey the laws of quantum mechanics or classical mechanics. As we now know, it would be impossible to quantize thermodynamics to achieve the smallest units of pressure, volume, temperature and entropy in order to arrive at statistical mechanics. Classical thermodynamics is not the $\hbar \to 0$ limit of quantum statistical mechanics. In fact, quantum mechanics is only used to count the number of states available at the microscopic level.

The real connection between thermodynamics and statistical mechanics is through the partition function. In statistical mechanics, the partition function of a system is defined as

$$Z = \sum_j e^{-\beta E_j} . \tag{1}$$

Here $\beta = 1/kT$, $k$ is the Boltzmann constant, and $E_j$ are the energy levels. The thermodynamic variables are then defined as:

Internal energy
$$U = -\frac{\partial \ln Z}{\partial \beta} , \tag{2}$$

Pressure
$$P = \frac{1}{\beta}\frac{\partial \ln Z}{\partial V} , \tag{3}$$

Entropy
$$S = k \ln Z + \frac{U}{T} . \tag{4}$$

We believe that gravity is in a similar situation. It would be impossible to quantize general relativity to achieve the smallest units of space and time in order to arrive at an underlying theory of gravity. A new structure with new degrees of freedom is needed. It is the statistical mechanics of spacetime.

**References**
[1]  G. 't Hooft, 1999 Nobel Lecture, *Rev. Mod. Phys*. **72** (2000) 333.
[2]  M. Veltman, 1999 Nobel Lecture, *Rev. Mod. Phys*. **72** (2000) 341.
[3]  R. Feynman, *Acta Physical Polonica* **24** (1963) 697.
[4]  B.S. DeWitt, *Phys. Rev. Lett*. **12** (1964) 742.
[5]  T. Rothman and S. Boughn, *Found. Phys*. **36** (2006) 1801.
[6]  A.A.Abdo et al., *Nature* **462** (2009) 331.
[7]  F. Tamburini et al., *Astron.Astrophys*. **A71** (2011) 553.
[8]  M. Eingorn and A. Zhuk, *Class. Quant. Grav*. **27**, 205014 (2010).
[9]  Y.K. Ha, *Int. J. Mod. Phys.: Conference Series* Vol. **7** (2012) 219.